\documentclass[aps,pra,reprint,superscriptaddress,twocolumn]{revtex4-1}
\usepackage[english]{babel}
\usepackage{ucs}
\usepackage[utf8x]{inputenc}
\usepackage{graphicx}
\usepackage{dcolumn}
\usepackage{bm}
\usepackage{bbm}
\usepackage{amsmath}
\usepackage{amssymb}
\usepackage{mathrsfs}
\usepackage{bbold}
\usepackage{empheq}
\usepackage{amsfonts}
\usepackage{color,colordvi}
\usepackage{physics}
\usepackage[colorlinks=true]{hyperref}
\usepackage[capitalize]{cleveref}
\usepackage{lipsum}

\begin{document}
\title{Loss asymmetries in quantum travelling wave parametric amplifiers}
\author{M.~Houde}
\affiliation{Department of Physics, McGill University, 3600 rue University, Montréal, Quebec H3A 2T8, Canada}
\author{L.~C.~G.~Govia}
\affiliation{Institute for Molecular Engineering, University of Chicago, 5640 S. Ellis Ave., Chicago, IL 60637}
\affiliation{Raytheon BBN Technologies, 10 Moulton St., Cambridge, MA 02138, USA}
\author{A.~A.~Clerk}
\affiliation{Institute for Molecular Engineering, University of Chicago, 5640 S. Ellis Ave., Chicago, IL 60637}

\begin{abstract}
We study theoretically how loss impacts the amplification and squeezing performance of a generic quantum travelling wave parametric amplifier.  Unlike previous studies, we analyze how having different levels of loss at signal and idler frequencies can dramatically alter properties compared to the case of frequency-independent loss.  We find that loss asymmetries increase the amplifier's added noise in comparison to the symmetric loss case.  More surprisingly, even small levels of loss asymmetry can completely destroy any quantum squeezing of symmetric collective output quadratures, while nonetheless leaving the output state strongly entangled.
\end{abstract}
\maketitle
\thispagestyle{plain}

\section{Introduction}


High fidelity qubit readout is a crucial ingredient to any viable quantum computing technology.  Superconducting qubits operated in circuit QED architectures are among the most promising platforms for quantum computing.  Here, readout is typically performed through homodyne detection of the output signal from a cavity dispersively coupled to a qubit \cite{Blais2004}.  High readout fidelity requires the use of quantum limited amplifiers \cite{Caves:1982aa,Clerk:2010aa} at the cavity output. The recently developed Josephson traveling-wave parametric amplifier (TWPA), a non-degenerate parametric amplifier built from a chain of Josephson junctions, offers this capability across a broad bandwidth of several GHz \cite{OBrien:2014aa,Macklin:2015aa,White:2015aa}, and will likely be a centerpiece of future circuit QED experiments of increasing size \cite{Fowler:2012aa,Versluis:2017aa}.

While TWPAs have proven to be excellent signal amplifiers, their utility is not limited to amplification. Even with only vacuum input, the output of an ideal TWPA exhibits
broadband two-mode squeezing and entanglement, and can be viewed as a source of two-mode squeezed vacuum states (TMSS).
Such states have a myriad of possible applications.  The collective symmetric quadratures of a TMSS are squeezed below vacuum, which directly enables enhanced readout protocols \cite{Barzanjeh2014,Didier:2015aa}.  Further, the signal-idler entanglement generated at the output could be used to entangle remote qubits  \cite{Kraus:2004aa,Didier:2017aa,Grimsmo:2017}, and opens up the possibility for many continuous variable protocols \cite{RevModPhys.77.513}, such as quantum teleportation \cite{Furusawa:1998aa}.

\begin{figure}[t]
  \includegraphics{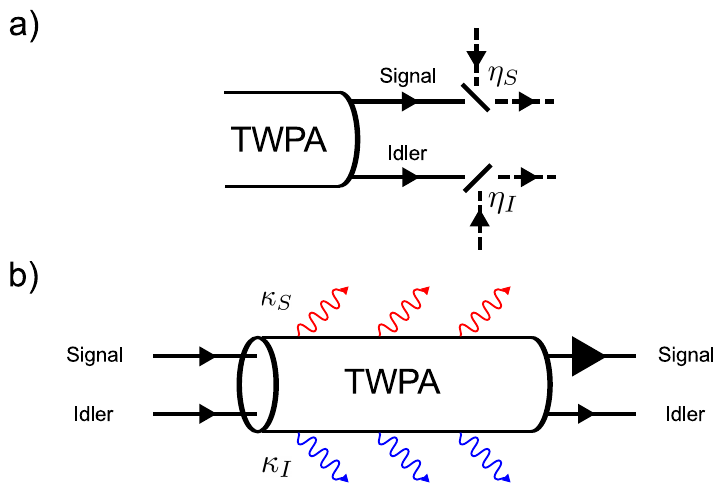}
  \caption{Schematic figure of two loss models. a) Beamsplitter loss model: signal and idler modes are passed through beamsplitters with transmission coefficients $\eta_{S}$ and $\eta_{I}$ respectively. b) Distributed loss model: signal and idler modes experience different decay rates $\kappa_{S}$ and $\kappa_{I}$, respectively, as they propagate through the TWPA.}
  \label{fig:setup}
\end{figure}

All of the above applications require high quality output states, implying that it is necessary to carefully model and understand how loss mechanisms in a TWPA degrade its output; experiments suggest such losses are non-negligible \cite{Macklin:2015aa}.  Previous theoretical studies of loss in TWPAs have either treated the loss as occurring only at the end of the device by introducing a fictitious beamsplitter \cite{Grimsmo:2017}, or have considered distributed loss throughout the device, but only for a degenerate amplification regime (where signal and idler are at the same frequency) \cite{Caves:87}. Furthermore, these previous treatments consider the case where signal and idler modes of the TWPA have equal (or symmetric) loss rates.

In this work, we extend the methodology of Refs.~\cite{Grimsmo:2017,Caves:87} to the more general situation of unequal (or asymmetric) signal and idler loss, in both the beamsplitter and distributed models of loss. Although we are motivated by recent work on Josephson TWPAs, our models and results are general, and thus also apply to more general TWPAs such as those of Refs.~\cite{HoEom:2012,Adamyan:2016,Vissers:2016aa,Erickson:2017}. Our key results concern the use of TWPAs as squeezing sources.  We find that asymmetric loss is detrimental to collective quadrature squeezing: asymmetry feeds amplified noise into the squeezed quadrature, thus quickly suppressing squeezing. This also has implications for applications to continuous variable teleportation \cite{Adesso:2005,Kim:2001}, as well as schemes for enhanced qubit readout \cite{Barzanjeh2014,Didier:2015aa}.

We further show how the output states of such models can be understood as thermal two-mode squeezed states.  This allows us to quantify the states' entanglement and purity in simple terms. With this description, we find that these quantities are hardly degraded by asymmetric loss. It is thus possible to have non-zero entanglement while having no squeezing of symmetric collective quadratures below the vacuum level  (see Fig.~\ref{fig:combinedplot}). For applications where one requires the symmetric quadratures to be squeezed, we propose a correction protocol to the lossy output state which allows one to regain squeezing below zero point.

We also analyze the impact of asymmetric loss on the amplification properties of a TWPA.  By analyzing the TWPA as a non-degenerate parametric amplifier with distributed loss, we find that asymmetry increases the level of gain (compared to the case of symmetric loss). The amount of noise added by the amplifier is also increased by loss asymmetries; nonetheless, the TWPA still remains nearly quantum limited.

%

This paper is organized as follows. In Sec.~\ref{sec:setup} we outline the generic model of a TWPA, and in Sec.~\ref{sec:th-TMSS} introduce an ideal way to describe the output of an imperfect TWPA as a thermal two-mode squeezed state.  In Sec.~\ref{sec:LE} we consider the effective beamsplitter model of a lossy TWPA including the effects of asymmetry; we do the same using the distributed model of loss in Sec.~\ref{sec:dist}.
We quantify the effect of symmetric and asymmetric distributed loss on the output squeezing and gain of the TWPA, thereby characterizing its use as both a squeezing source and an amplifier. Further, we explore the interplay between phase mismatch \cite{OBrien:2014aa} and asymmetric distributed loss. Our conclusions are finally presented in Sec.~\ref{sec:conc}.




\section{Model of an ideal TWPA}
\label{sec:setup}


As is standard \cite{Caves:87}, we model a generic TWPA as a nonlinear one-dimensional transmission line or waveguide.  The basic amplification process involves driving the system with a large coherent pump tone (frequency $\bar{\omega}_P$, wavevector $k_P >0$), which is then scattered by the nonlinearity into photons at signal and idler frequencies.
For a four-wave mixing nonlinearity (as is relevant to setups employing Josephson junctions), two pump photons are converted to a pair of signal and idler photons.  We will consider signal photons with frequency $\omega_{S}\in (\bar{\omega}_{S}-D,\bar{\omega}_{S}+D)$, where $D$ is the maximum bandwidth of interest.   Energy conservation then determines the relevant range of idler photons via $\omega_{I}=2 \bar{\omega}_{P}-\omega_S$.   We will focus exclusively on non-degenerate modes of operation, where signal and idler frequencies do not overlap;
we thus take $\bar{\omega}_S-D > \bar{\omega}_P$.  Given this condition, we can treat signal and idler photons (in the bandwidth of interest) as propagating in effectively independent one-dimensional bosonic channels.  We take these fields to have infinite extent, with a non-zero interaction only between positions $x=0$ and $x=L$.  Further, only right-moving signal and idler fields will be phase-matched to the pump, hence we do not consider left-moving fields (which are decoupled from the dynamics).

To write the system Hamiltonian, we will work in a rotating frame at frequency $\bar{\omega}_S$ ($\bar{\omega}_I = 2 \bar{\omega}_P - \bar{\omega}_S$) for the signal (idler) channel.
Treating the nonlinear interaction at a mean-field level and setting $\hbar = 1$, the basic TWPA Hamiltonian is:
\begin{align}\label{eqn:HamDL}
  \hat{H} = \int dx\hspace{3pt}
  	\Bigg[ &\hat{a}^{\dagger}_{S}(x)\left( -iv_S \partial_{x} \right)\hat{a}_{S}(x) +\hat{a}^{\dagger}_{I}(x)\left( -iv_I \partial_{x} \right)\hat{a}_{I}(x)\nonumber\\
  &+\frac{i}{2}\left( \nu(x)\hat{a}^{\dagger}_{S}(x)\hat{a}^{\dagger}_{I}(x) -h.c.\right) \Bigg],
\end{align}
Here, $v_n$ denotes the group velocity of channel $n = S,P,I$, and the lowering operator for channel $n$ is $\hat{b}_{n}(x) = e^{i k_n x} \hat{a}_n(x)$, where $k_{n}=\bar{\omega}_{n}/v_{n}$.  The operators $\hat{a}_n(x)$ describe the spatial envelope of the signal and idler fields, and are canonical bosonic fields:  $\left[ \hat{a}_{n}(x),\hat{a}^{\dagger}_{m}(x')   \right]=\delta(x-x')\delta_{nm}$.   Finally,
 the parametric interaction, with units of a rate, is
 \begin{align}\label{eqn:parampint}
   \nu(x)= \lambda  |\psi_{P}|^{2} e^{i(2k_{P}-k_{S}-k_{I})x}\left[ \theta(x)-\theta(x-L) \right],
 \end{align}
 where $\psi_{P}$ is the classical pump amplitude, $\lambda$ is the bare four-wave mixing interaction strength, $\theta(x)$ is the Heaviside function, and the exponential factor accounts for a lack of phase matching.  We start by assuming all group velocities are the same, implying perfect phase matching;
$\nu(x)$ can then be taken to be real and positive without any loss of generality.

Working in the Heisenberg picture, the output of our amplifier is described by the operators $\hat{a}_{S/I}(L,t)$, and the input by $\hat{a}_{S/I}(0,t)$.  For the ideal TWPA described by Eq.~(\ref{eqn:HamDL}), one can easily solve the Heisenberg equations of motion for the system.  By relating output fields to input fields in the frequency domain, one finds the basic scattering relations that characterize the system as a non-degenerate (phase-preserving) amplifier:
\begin{align}
  \hat{a}_{S}[L,\omega] =&
  	e^{i\omega L/v}\sqrt{G_{\rm ideal}} \hat{a}_{S}[0,\omega]\nonumber\\
		&+ e^{i\omega L/v}\sqrt{G_{\rm ideal}-1}  \hat{a}^{\dagger}_{I}[0,\omega],
\end{align}
where we have taken the Fourier transform of our fields, and defined the power gain as
\begin{align}\label{eqn:idealgain}
  G_{\rm ideal} = \cosh^2(L \nu / v) \equiv \cosh^{2}(r),
\end{align}
with $r$ denoting the frequency-independent squeezing parameter for our model.  Note that in a more realistic model, $r$ will be frequency dependent due to e.g.~dispersion effects that cause a lack of phase matching.

In the case where the input fields are just vacuum noise, the output of the ideal TWPA is characterized by the correlation functions
\begin{align}
  &\left<\hat{a}^{\dagger}_{S/I}[L,\omega]\hat{a}_{S/I}[L,\omega']\right> = 2\pi \sinh^2(r)\delta(\omega+\omega'), \label{eqn:idnum} \\
  &\left<\hat{a}_{S}[L,\omega]\hat{a}_{I}[L,\omega']\right> = \pi\sinh(2r)\delta(\omega+\omega'). \label{eqn:idanom}
\end{align}
Note that the frequency-conserving delta functions imply that there are no $L$-dependent phase factors above. These correlators imply we have perfect two-mode squeezing at each frequency.

Introducing Hermitian quadrature operators via $\hat{a}_{S/I}(L,t) = (\hat{X}_{S/I}(L,t) + i \hat{P}_{S/I}(L,t))/\sqrt{2}$, we define the
symmetric collective quadratures
\begin{align}
	\hat{X}_{\pm}(L,t) = \frac{\hat{X}_S(L,t) \pm \hat{X}_I(L,t)}{\sqrt{2}}, \label{eqn:pmout}
\end{align}
with a similar definition for $\hat{P}_{\pm}(L,t) $.
The noise spectral density of a generic quadrature is defined as:
\begin{align}
  S_{\hat{X}}[\omega]= \frac{1}{2} \int^{\infty}_{-\infty} dt \hspace{3pt}e^{i \omega t}\left< \left\{\hat{X}(t), \hat{X}(0)  \right\}\right>.
\end{align}

One finds that with the choice of interaction phase in Eq.~(\ref{eqn:HamDL}),  the ideal TWPA squeezes fluctuations in both the $\hat{X}_-$ and $\hat{P}_+$ quadratures.  The noise spectral density of these squeezed quadratures are
\begin{align}\label{eqn:idealtwpa}
 	S^{\rm ideal}_{\hat{X}_{-}}[\omega] = S^{\rm ideal}_{\hat{P}_{+}}[\omega]
	=\frac{1}{2}e^{-2r}.
\end{align}
For any $r>0$ we obtain squeezing below zero point ($S_{X_{-}}[\omega] =1/2$).




\section{Thermal two-mode squeezed state parameterization of a lossy TWPA}
\label{sec:th-TMSS}

The main goal of this paper is to characterize the output state of an imperfect TWPA.  Losses will degrade the perfect two-mode squeezing of signal and idler generated at each frequency by an ideal TWPA.  Generically, the state at each frequency will now be a thermal two-mode squeezed state (th-TMSS).  Such a state has the form
\begin{align}
	\label{eq:thTMSSrho}
  \hat{\rho}_{\rm th-TMSS} = \hat{S}_{2}(R)\left[\hat{\rho}^{\rm th}_{S}(\bar{n}_{S})\otimes\hat{\rho}^{\rm th}_{I}(\bar{n}_{I})     \right]\hat{S}^{\dagger}_2(R),
\end{align}
where $\hat{S}_2(R) = \exp\left[R\left(\hat{B}_{S}\hat{B}_{I} - h.c.\right)\right]$ is the two-mode squeezing operator for bosonic modes $\hat{B}_{S/I}$ with squeezing parameter $R$, and $\hat{\rho}^{\rm th}_{i}(\bar{n}_{i})$ describes a single-mode thermal state with average photon number $\bar{n}_{i}$. An imperfect TMSS can be fully described by these three parameters: $\bar{n}_{S},\bar{n}_{I}$, and $R$. In general $R$ can be complex, however, we can always work in a gauge where $R$ is real.

A general th-TMSS has the following non-zero correlators:
\begin{align}
  \left<\hat{B}_{i}^\dagger\hat{B}_{i}\right>
  &= \bar{n}_i + (\bar{n}_S + \bar{n}_I+1)\sinh^2(R), \label{eqn:THnum} \\
  \left<\hat{B}_{S}\hat{B}_{I}\right>
&= \frac{\bar{n}_S + \bar{n}_I +1}{2}\sinh(2R),\label{eqn:THanom}
\end{align}
where $i=S,I$.
Note that $\bar{n}_S - \bar{n}_I$ determines the asymmetry of the state (e.g.~how different is the state if we exchange the $S$ and $I$ modes). The anomalous entanglement properties of asymmetric th-TMSS  have been discussed in Ref.~\cite{Adesso2004}.


Important properties of the state have a simple expression in terms of the th-TMSS description.  The purity $\mu$ of the state is independent of $R$ and is given by
\begin{equation}
 	\mu	\equiv \mathrm{ Tr } \left(\hat{\rho}_{\rm th-TMSS}\right)^2
		=\frac{1}{(1+2\bar{n}_{S})(1+2\bar{n}_{I})}.
\end{equation}
The entanglement of the modes $S$ and $I$ in this state can be characterized by the logarithmic negativity \cite{RevModPhys.77.513}, and takes the form \cite{Wang:2015aa}
\begin{align}\label{eqn:logneg}
 E_{\rm N} = -\ln\left[n_{R}-\sqrt{n_{R}^{2}-(1+2\bar{n}_{S})(1+2\bar{n}_{I})}\right],
\end{align}
where
\begin{align}
 n_{R}=(\bar{n}_{S}+\bar{n}_{I}+1)\cosh(2R).
\end{align}

Introducing Hermitian quadrature operators via $\hat{B}_{S/I}=(\hat{X}_{S/I}+i\hat{P}_{S/I})/\sqrt{2}$, we define symmetric collective quadratures $\hat{X}_{\pm}=(\hat{X}_{S}\pm\hat{X}_{I})/\sqrt{2}$.  A crucial quantity is the variance of the squeezed collective quadrature  $\hat{X}_{-}$,
\begin{align}
S_{\hat{X}_{-}}& \equiv \langle \hat{X}^{2}_{-}  \rangle -\langle \hat{X}_{-}  \rangle^2\
 =\frac{1}{2}\left[1+\bar{n}_S + \bar{n}_I\right]e^{-2R}.
\end{align}

Finally, note that the multimode output of the TWPA (even with loss) can be understood as a product of th-TMSS states.
For each frequency $\omega$ of interest, we can introduce frequency-resolved temporal modes, defined as
\begin{align}
  \hat{B}^{\rm out}_{S/I}[\omega]=\lim_{\delta \to 0}\frac{1}{\sqrt{\delta }}
  	\int_{-\delta /2}^{\delta /2}d\omega' \, \hat{a}^{\rm out}_{S/I}[\pm \omega + \omega'],
  \label{eq:TempModes}
\end{align}
where the $+$ ($-$) sign is for the signal (idler) mode.  These modes have center frequency $\pm \omega$, a vanishing bandwidth $\delta $, and satisfy $[ \hat{B}_j[\omega], \hat{B}_{j'}^\dagger[\omega] ] = \delta_{j,j'}$.  For each frequency $\omega$, the pair of modes $ \hat{B}^{\rm out}_{S/I}[\omega]$ will be a th-TMSS state of the form in
Eq.~(\ref{eq:thTMSSrho}), and thus can be completely parameterized by $\bar{n}_S, \bar{n}_I, R$.

%
%
%

\section{Lumped element loss}
\label{sec:LE}

We begin our treatment of loss in a TWPA by considering the simplest possible model of loss, the so-called ``lumped element model".  Here, we model the final output of a lossy TWPA by applying an independent beamsplitter transformation to each output ($S$,$I$) of an ideal TWPA, see Fig.~\ref{fig:setup}a).  The non-unity transmission of the beamsplitters corresponds to loss.  This model is described by the transformation
\begin{align}\label{eqn:bs1}
	&\hat{a}^{\rm out}_{S}[\omega] = \sqrt{\eta_S[\omega]}\hat{a}_{S}[L,\omega] + \sqrt{1-\eta_S[\omega]}\hat{\xi}_{S}[\omega], \\
	&\hat{a}^{{\rm out}}_{I}[\omega]  = \sqrt{\eta_I[\omega]}\hat{a}_{I}[L,\omega] + \sqrt{1-\eta_I[\omega]}\hat{\xi}_{I}[\omega], \label{eqn:bs2}
\end{align}
where $\hat{a}_{S/I}[L,\omega]$ are the modes leaving the amplification region of the ideal TWPA, $\eta_{S/I}[\omega]$ are the transmission rates of the signal/idler through the beamsplitters,
and $\hat{\xi}_{S/I}[\omega]$ are the noise modes coming from the other input ports of the beamsplitters. We take this noise to be simple delta-correlated vacuum noise.
Using Eqs.~\eqref{eqn:bs1} and \eqref{eqn:bs2}, we find the output field variances that characterize the output of the lossy TWPA:
\begin{align}
  &\left<\left[\hat{a}^{{\rm out}}_{S}[\omega']\right]^\dagger\hat{a}^{\rm out}_{S}[\omega]\right> = 2\pi\eta_{S}[\omega] \sinh^2(r)\delta(\omega+\omega'), \label{eqn:LEnum} \\
  &\left<\left[\hat{a}^{{\rm out}}_{I}[-\omega']\right]^\dagger\hat{a}^{\rm out}_{I}[-\omega]\right> = 2\pi\eta_{I}[-\omega] \sinh^2(r)\delta(\omega+\omega'), \\
    &\left<\hat{a}^{\rm out}_{S}[\omega]\hat{a}^{\rm out}_{I}[\omega']\right> =
  	\pi\sqrt{\eta_S[\omega] \eta_I[\omega'] }\sinh(2r)\delta(\omega+\omega'). \label{eqn:LEanom}
\end{align}

Recall that the signal and idler channels correspond to different frequency intervals of the single nonlinear transmission line that makes up the TWPA.  In the original lab frame,
$\eta_{S}[\omega]$ describes loss at frequency $\bar{\omega}_S + \omega$, whereas $\eta_{I}[\omega]$ describes loss at frequency $\bar{\omega}_I + \omega = 2 \bar{\omega}_P - \bar{\omega}_S + \omega$.  It is thus entirely possible that these channels will experience different levels of loss.  The simplest way to model this is to allow the transmissions $\eta_S[\omega]$ and $\eta_I[\omega]$ to differ from one another.  While the effects of symmetric lumped element loss, $\eta_S = \eta_I$, have been studied previously in Ref.~\cite{Grimsmo:2017}, asymmetric loss effects have not.

To quantify the effects of loss asymmetry on the two-mode squeezing between $\hat{a}_S[\omega]$ and $\hat{a}_I[-\omega]$, we will use the following parameterization:
\begin{align}\label{eqn:eta}
   \eta_{S}[\omega]=1-\bar{\epsilon}(1 + \delta),
   \hspace{0.75 cm} \eta_{I}[-\omega]=1-\bar{\epsilon}(1- \delta).
\end{align}
Here $\bar{\epsilon}$ describes the average loss and $\delta$ is the relative asymmetry;
we have suppressed the explicit $\omega$ dependence of $\epsilon, \delta$.  Without loss of generality we take $\delta > 0$, implying that the signal mode has higher loss than the idler mode.
 As discussed in detail below, we find that asymmetry in the loss (i.e.~non-zero $\delta$) starts to play a significant role when the
 average amount of loss is large enough to disrupt the squeezing of an ideal TWPA.
 This corresponds to the condition  $\bar{\epsilon} \gtrsim e^{-2r}$ (c.f. Eq.~(\ref{eqn:idealtwpa})).



\subsection{Thermal two-mode squeezed states (th-TMSS)}

As discussed in Sec.~\ref{sec:th-TMSS}, for each frequency $\omega$, the output of the lossy TWPA can be mapped onto a thermal TMSS using Eqs.~(\ref{eq:TempModes}) and
Eqs.~(\ref{eqn:THnum}),(\ref{eqn:THanom}).  We will use this parameterization to discuss the effect of loss.

\subsubsection{Weak average loss}
Consider first the limit of weak average loss, $\bar{\epsilon} \ll 1$ and a large intrinsic squeeze parameter $r$.  Useful expressions are obtained by taking the large-$r$ and small $\bar{\epsilon}$ limit while keeping $\bar{\epsilon}e^{2r}$ finite and small. This amounts to an expansion in $\bar{\epsilon}e^{2r}$.  For purely symmetric loss, $\bar{\epsilon}e^{2r} \ll 1$ implies the amount of vacuum noise added from the loss channels to the output is small enough to not  appreciably change the squeezing in the output.


Following this procedure, the effective thermal occupancies $\bar{n}_{S/I}$ are given to second order in $\bar{\epsilon}$ by
\begin{align}
  \bar{n}_{S/I} \approx
  	\frac{1}{4} \left(\bar{\epsilon} e^{2r} \right) (1\pm \delta)
	- \frac{1}{16} \left(\bar{\epsilon} e^{2r} \right)^2 (1- \delta^{2})
  + \mathcal{O}(\bar{\epsilon}^3), \label{eqn:smallepsN}
\end{align}
where $+/-$ corresponds to $S/I$.  Similarly, the effective squeezing parameter is given by:
\begin{align}
  \frac{\cosh{(R)}}{\cosh{(r)}} \approx
  	1- \frac{1}{4} \left( \bar{\epsilon} e^{2r}\right) +
		\frac{1}{32}\left( \bar{\epsilon} e^{2r}\right)^2 \left(5-2\delta^{2} \right)  + \mathcal{O}(\bar{\epsilon}^3).
\end{align}
Thus, in this regime the effect of loss asymmetry is minimal:  it only changes the coefficients in the expansions for each parameter, and is not exponentially enhanced (compared to the symmetric loss case).  As we now show, this is not true for larger levels of loss.
%
%
%
%
%
%


\subsubsection{Larger average loss, weak asymmetry}
For larger values of average loss, we consider the large-$r$ limit where $\bar{\epsilon}$ is no longer arbitrarily small. In this case, $\bar{\epsilon}e^{2r}$ is no longer a small parameter and we therefore cannot expand with respect to it. Insight is instead obtained by first assuming a weak asymmetry ($\delta\ll1$) and expanding in $\delta$.
 We again consider the case of large intrinsic squeezing, and take the asymptotic large-$r$ form of each coefficient in our expansion while keeping $\bar{\epsilon}$ fixed.  Doing this, the effective thermal numbers are given by
\begin{align}\label{eqn:thermalpop}
 \nonumber\bar{n}_{S/I}\approx &\frac{\sqrt{(1-\bar{\epsilon})\bar{\epsilon}}}{2} e^{r} \pm \frac{e^{2r}}{4}\bar{\epsilon} \delta + \frac{1}{16\sqrt{(1-\bar{\epsilon})\bar{\epsilon}}} e^{3r}\bar{\epsilon}^{2}\delta^{2} \\ &+ \mathcal{O}(\delta^3).
\end{align}
We now see that asymmetry has a dramatic effect:  for $\delta = 0$ the thermal numbers scale as $e^r$, whereas with asymmetry (i.e.~$\delta \neq 0$), there is a much stronger heating scaling as $e^{3r}$.

For the effective squeezing parameter in the same weak-$\delta$, large-$r$ regime, we find
\begin{align}\label{eqn:Reff}
 \nonumber\frac{\cosh(R)}{\cosh(r)}\approx&
 	\frac{\sqrt{1-\bar{\epsilon }}}{\left(1+(1-\bar{\epsilon})\bar{\epsilon}e^{2r}\right)^{1/4}} \\
	&\times\left[
		1 - \frac{ \bar{\epsilon}^{2}\delta^{2}e^{4r}  }{16\left[ 1+(1-\bar{\epsilon})\bar{\epsilon}e^{2r}    \right]}
	+ \mathcal{O}(\delta^3) \right].
\end{align}
Again, we see that the loss-induced suppression of $R$ is more pronounced in the asymmetric loss case.

\begin{figure}
  \includegraphics[width=\linewidth]{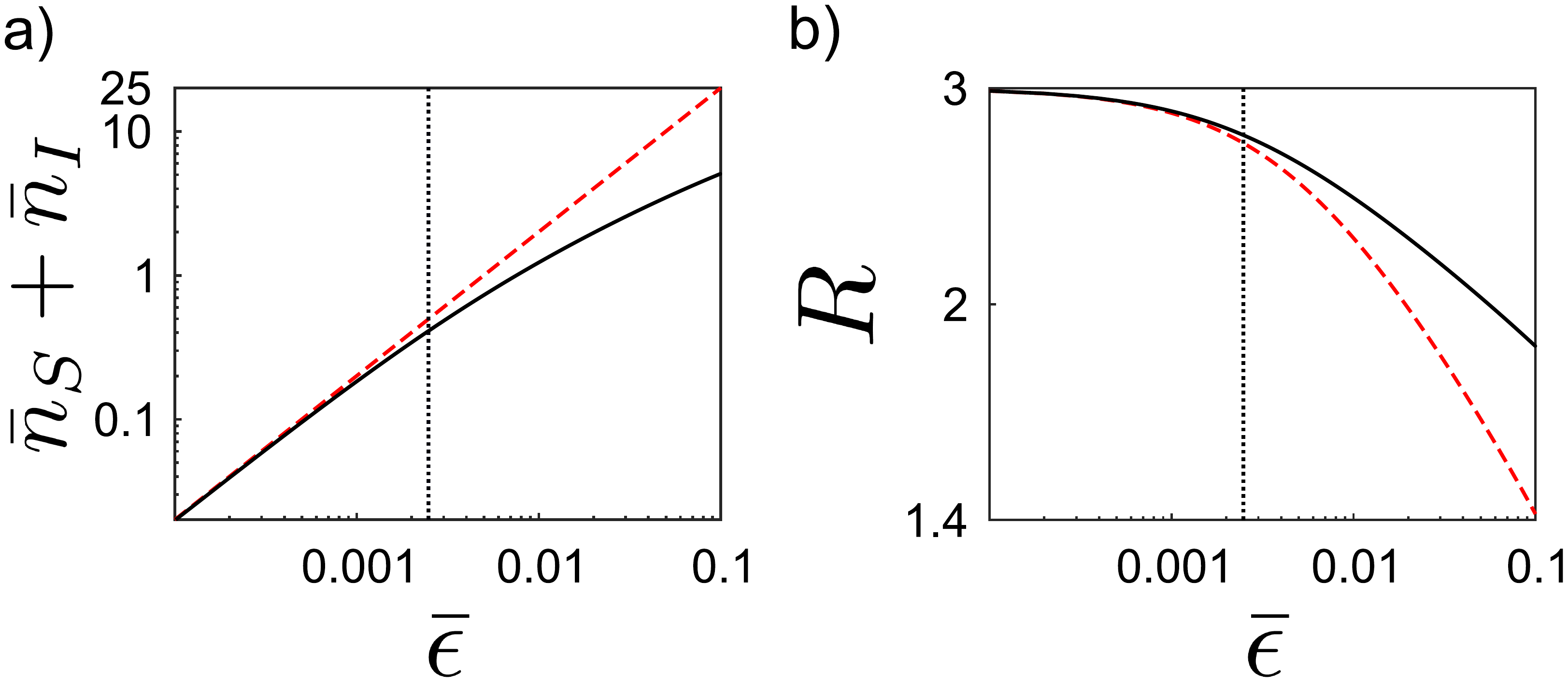}
  \caption{Properties of the TWPA output state as a function of average loss for the effective beamsplitter model (c.f. Sec. \ref{sec:LE}). Using the th-TMSS description, we plot a) the sum of the average effective thermal photon populations ($\bar{n}_{S} + \bar{n}_{I}$) and b) the effective squeezing parameter $R$ for symmetric (black-solid curve) and fully asymmetric (red-dashed curve, asymmetry parameter $\delta=1$) loss, as functions of the average loss $\bar{\epsilon}$.  The asymmetry parameter is defined in Eq.~(\ref{eqn:eta}). The vertical dotted lines are markers for the crossover point, $\bar{\epsilon} = e^{-2r}$; for larger $\bar{\epsilon}$ loss asymmetry has a strong impact.  The ideal TWPA output squeezing parameter is $r=3$ for both figures, corresponding to a gain of $20 $dB in the loss-free case.}
  \label{fig:thermal}
\end{figure}

The above analysis suggests the existence of a kind of crossover:  for weak average loss $\bar{\epsilon} < e^{-2r}$, loss asymmetry has a minor effect on our output state, whereas for larger $\bar{\epsilon}$ it has a pronounced effect.  This behaviour is highlighted in Fig.~\ref{fig:thermal}, where we compare the exact behaviour of $\bar{n}_{S/I}$ and $R$ as a function of average loss, with and without asymmetry; the crossover scale  $\bar{\epsilon} \sim e^{-2r}$ beyond which asymmetry is important is clearly seen.

\subsection{Squeezing below vacuum}



\subsubsection{Symmetric loss}

Recall that our choice of pump phase ensures that without loss, the collective $\hat{X}_-$ and $\hat{P}_+$ quadratures will be squeezed, c.f.~Eq.~(\ref{eqn:pmout}).
For the symmetric loss case ($\eta_{S}=\eta_{I}\equiv\eta$), we find directly from Eqs.~\eqref{eqn:LEnum} and \eqref{eqn:LEanom} that
\begin{align}
S^{\rm sym}_{\hat{X}_{-}}[\omega]&=\frac{1}{2}\left[1-\eta+\eta e^{-2r}\right], \label{eqn:symLE}
\end{align}
with a similar result for $S^{\rm sym}_{\hat{P}_{+}}[\omega]$. As might be expected, with symmetric loss, we simply interpolate between the perfect squeezed state at $\eta=1$ and a vacuum state when $\eta=0$.  Note that unless the loss level is $100\%$, there is always some squeezing in the output in this symmetric loss case.


\subsubsection{Asymmetric loss}

\begin{figure}
  \includegraphics[width=0.8\columnwidth]{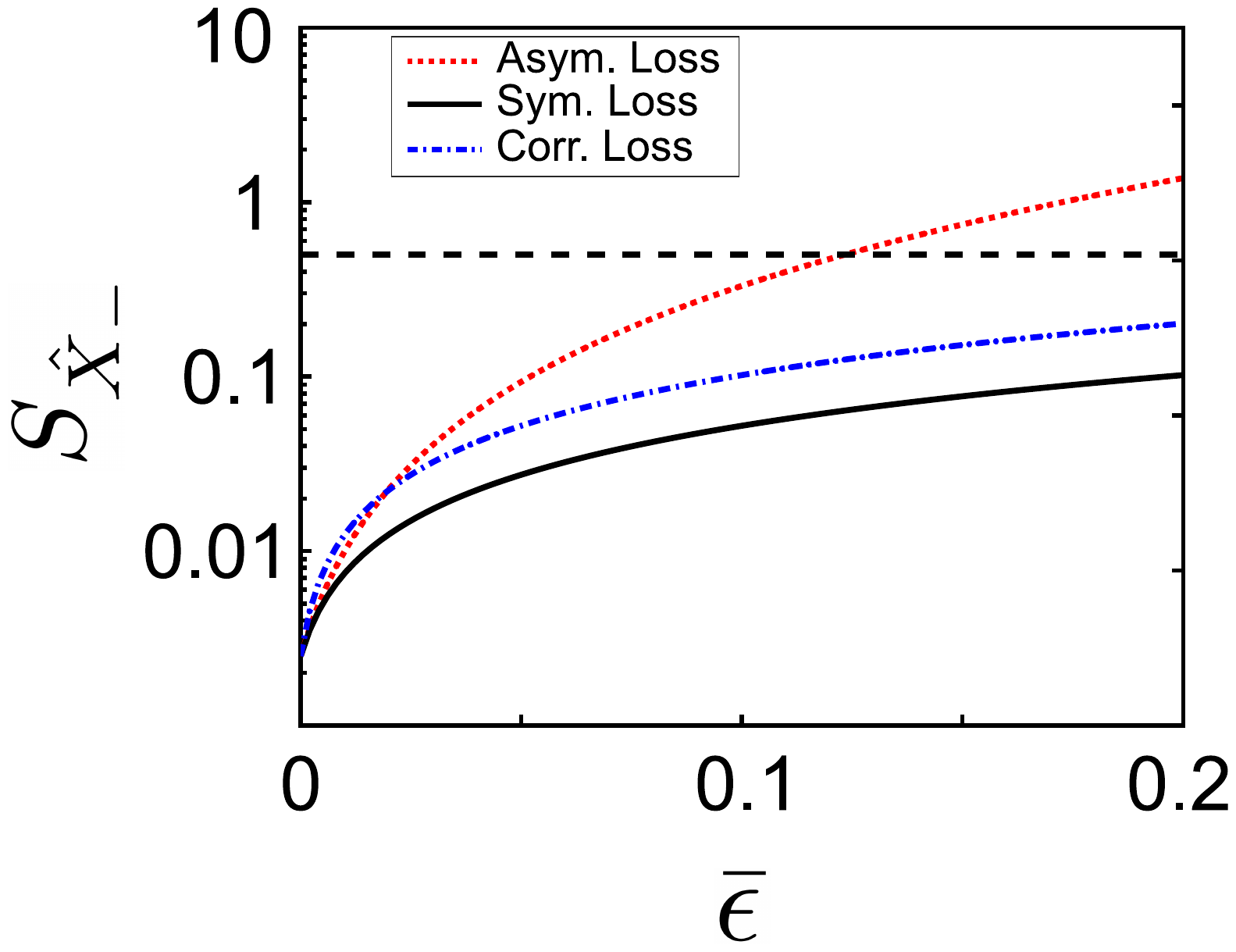}
  \caption{Output squeezing of the output state of a TWPA, for the beamsplitter model (c.f. section \ref{sec:LE}), as a function of average loss $\bar{\epsilon}$. The squeezing parameter $r = 2.65$, corresponding to a gain of $\sim 17$ dB. For a fully asymmetric situation where only the signal mode is lossy (i.e.~$\delta = 1$ in Eq.~(\ref{eqn:eta}), red curve), loss can destroy any squeezing below the vacuum level.  In contrast,  if the loss is symmetric, one always has squeezing below vacuum (black curve).  Even with fully asymmetric loss, one can use our proposed correction scheme (c.f. section \ref{sec:correction}) to regain squeezing below zero-point (blue curve).  Note that while asymmetric loss can kill vacuum squeezing, signal-idler entanglement always remains non-zero (see Fig.~\ref{fig:logplot}).}
  \label{fig:combinedplot}
\end{figure}

When the transmission rates for the signal and idler modes are different, the noise spectral density of the minus quadrature of the output field is
\begin{align} \label{eqn:outvar}
	S^{\rm asym}_{\hat{X}_{-}}[\omega]   =
		\frac{1}{2}
			&\left[ \left(1-\frac{\eta_S+\eta_I}{2} \right)+\frac{e^{-2r}}{4}\left(\sqrt{\eta_S}+\sqrt{\eta_I}\right)^2 \right.\nonumber\\
		&\left.+ \frac{e^{2r}}{4}\left(\sqrt{\eta_S}-\sqrt{\eta_I}\right)^2\right].
\end{align}
In this expression, $\eta_S$ ($\eta_I$) is evaluated at frequency $+\omega$ ($-\omega$). The first bracketed term of Eq.~\eqref{eqn:outvar} describes the vacuum noise added to the output field as a result of the lossy transmission lines, and the second term describes the usual squeezed noise (which is suppressed when $\eta_s, \eta_I < 1$).  The third term, unique to asymmetric loss, describes amplified noise
$\propto e^{2r}$  that is now mixed into the minus quadrature due to the asymmetry in the beamsplitters' transmission rates.  This mixing in of amplified noise is clearly detrimental to achieving squeezing below zero point. In Fig.~\ref{fig:combinedplot} we see that for larger levels of average loss, the squeezing for asymmetric loss is above zero point, whereas for symmetric loss it is still below zero point. Loss asymmetries can thus have a large impact on the production of squeezing, and will greatly affect schemes that use the output of a TWPA as a squeezing source, such as two-mode qubit readout \cite{Didier:2015aa} and continuous variable teleportation \cite{Furusawa:1998aa,Adesso:2005,Kim:2001}.


\subsubsection{Squeezing of asymmetric collective quadratures}


While the symmetric $\hat{X}_{-}$ collective quadrature rapidly becomes unsqueezed with loss asymmetry, one might ask whether there are other collective quadratures that remain squeezed.  It is easy to verify that any symmetric collective quadrature of the form $\sqrt{2}\hat{X}_{\rm sym} = \hat{X}_S + e^{i\phi}\hat{X}_I$ will have a contribution from amplified noise ($\propto e^{2r}$) in its noise spectral density when there is loss asymmetry; hence, loss asymmetry prevents any such quadrature from being squeezed.

That being said, one can define {\it asymmetric} collective quadratures (i.e.~$S$ and $I$ modes weighted unequally) that exhibits squeezing even with asymmetric loss.  We define
\begin{align}
	\hat{X}_-^{\rm asym} &= \cos\theta \hat{X}^{\rm out}_S - \sin\theta \hat{X}^{\rm out}_I, \\
	\hat{P}_+^{\rm asym} & = \cos\theta \hat{P}^{\rm out}_S + \sin\theta \hat{P}^{\rm out}_I.
\end{align}
By taking the parameter $\tan\theta =\sqrt{\eta_S/\eta_I}$, one finds
\begin{align}
	S_{\hat{X}^{\rm asym}_-}[\omega]= \frac{1}{2}\left[1 - \frac{2\eta_S\eta_I}{\eta_S+\eta_I}\left(1- e^{-2r}\right)\right],
\end{align}
with a similar result for $S_{\hat{P}^{\rm asym}_+}$; again, in this expression $\eta_S$ ($\eta_I$) are evaluated at frequency $+\omega$ ($-\omega$).  We thus see that these quadratures are squeezed below vacuum whenever $r>0$, irrespective of loss asymmetries.  Note crucially that the definition of this quadrature depends sensitively on the amount of loss asymmetry; further for $\theta \neq \pi/4$, the squeezed collective quadratures
$\hat{X}^{\rm asym}_-,\hat{P}^{\rm asym}_+$ do not commute with one another.

The utility of having such non-commuting, asymmetric quadratures squeezed is mixed.  They do imply the presence of entanglement, as they allow violation of generalized versions of the well known Duan and Tan inequalities \cite{Tan:1999,RevModPhys.77.513}.  As we will see in Sec.~\ref{sec:PurLN}, this implies that loss asymmetry does not prevent using the TWPA output state to entangle other systems.  However, there are other applications that crucially require two commuting joint quadratures to be squeezed, e.g.~the enhanced dispersive measurement scheme described in Ref.~\cite{Didier:2015aa}.




\subsection{Purity and logarithmic negativity}
\label{sec:PurLN}

We now study how loss (modelled using the lumped-element approach) impacts the purity and entanglement (as measured by the logarithmic negativity \cite{RevModPhys.77.513}) of the TWPA output state at a given frequency.

\subsubsection{Symmetric loss}

Without any loss asymmetry (i.e.~$\delta = 0$ in Eq.~(\ref{eqn:eta})), the log negativity is given by:
\begin{align}
    E_{\rm N}= - \ln\left[e^{-2r}+(1-e^{-2r})\bar{\epsilon}\right].
\end{align}
This saturates to $E_{\rm N}=  \ln\left[1/\bar{\epsilon}\right]>0$ in the large $r$ limit. The logarithmic negativity mimics the behaviour of the symmetric squeezed quadrature (see Eq.~(\ref{eqn:symLE})): it decreases monotonically from $2 r$ to $0$ as the loss $\bar{\epsilon}$ increases.

In contrast, the purity of the state for symmetric losses is given by
\begin{align}
  \mu= \frac{1}{1+2(1-\bar{\epsilon})\bar{\epsilon}(\cosh(2r)-1)}.
\end{align}
For any nonzero loss $\bar{\epsilon}$, the purity decays exponentially as $e^{-2r}$ in the large-$r$ limit.
Thus, for large intrinsic squeezing $r$, even a small amount of loss leads to
a highly impure output state that nonetheless possesses a potentially large logarithmic negativity.  The utility of such a state in potential applications is thus at first glance somewhat suspect.

\begin{figure}[t]
  \includegraphics[width=0.764\columnwidth]{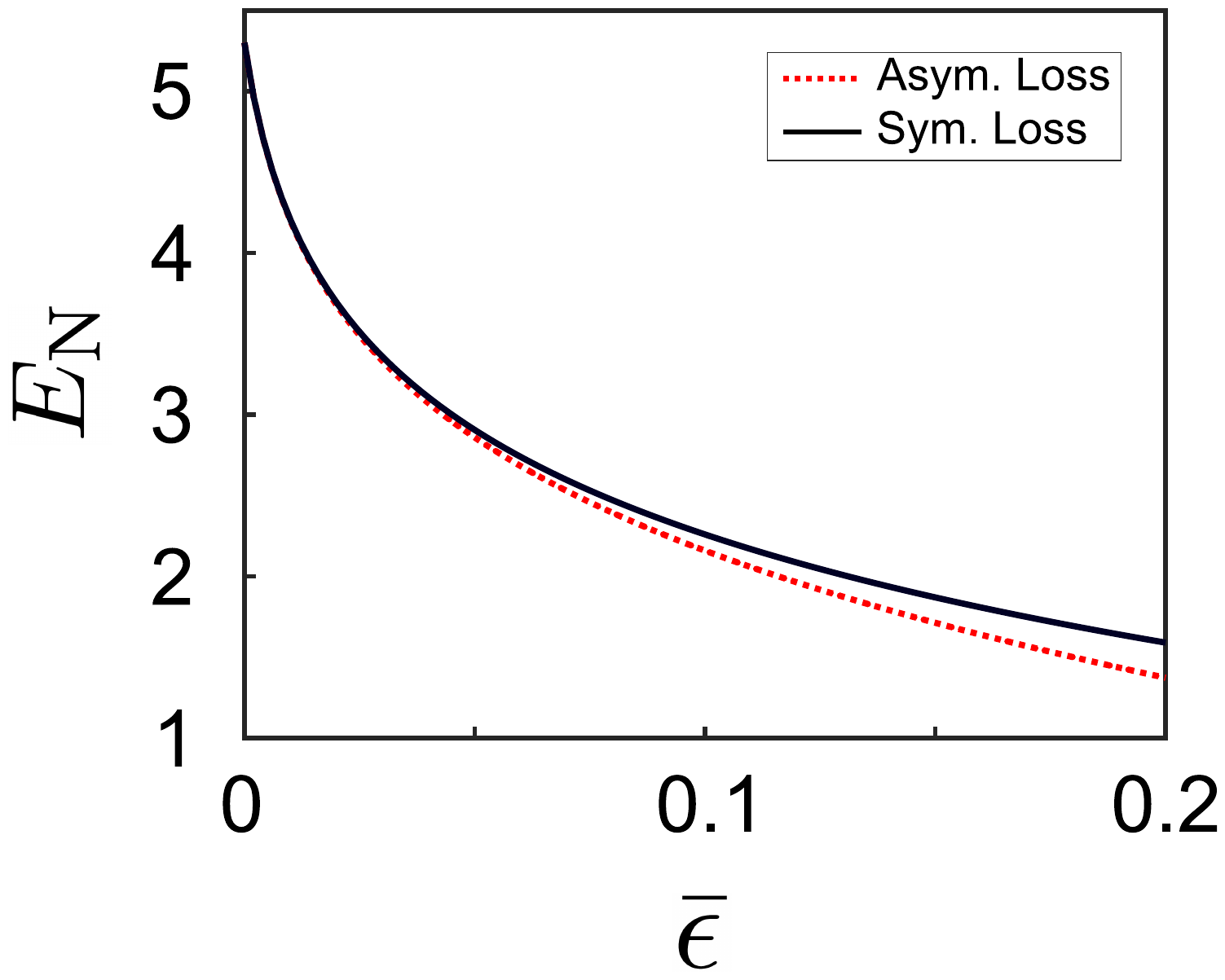}
  \caption{Logarithmic negativity of the output state of a TWPA, for the beamsplitter model (c.f. section \ref{sec:LE}), as a function of average loss $\bar{\epsilon}$. The squeezing parameter $r = 2.65$, corresponding to a gain of $\sim 17$ dB. For asymmetric loss (red curve), we consider the maximally asymmetric case where all loss is in the signal mode (i.e. $\delta=1$ in Eq.~(\ref{eqn:eta})). Logarithmic negativity is much less sensitive to asymmetry (c.f. section \ref{sec:PurLN}). Even with asymmetry we retain non-zero logarithmic negativity.}
  \label{fig:logplot}
\end{figure}

To test the utility of such an entangled thermal two-mode squeezed state, we consider the remote entanglement protocol of Ref.~\cite{Kraus:2004aa}.  Here, the signal and idler of a TMSS are each sent to a separate qubit, with the goal of stabilizing a two-qubit entangled state (see Appendix \ref{app:Qubit} for further details). In the ideal (zero loss) case, when the signal (idler) qubit is resonant with the signal (idler) mode, the steady-state of the two-qubit system is a pure entangled state, and reaches a maximally entangled Bell state in the large gain limit \cite{Kraus:2004aa}.

\begin{figure}[t]
  \includegraphics[width=0.75\columnwidth]{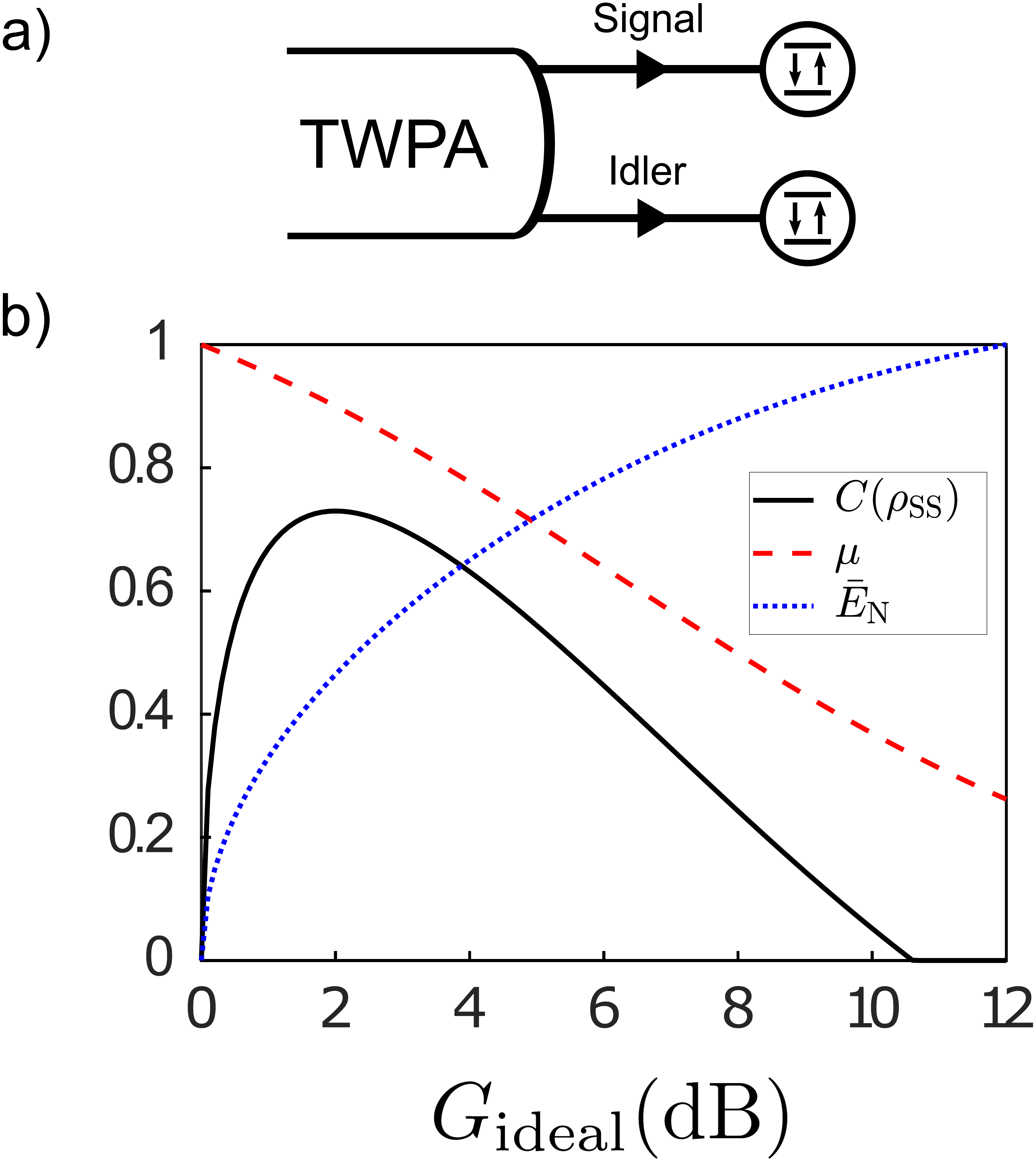}
  \caption{Signal and idler modes at the output of a TWPA are each sent to separate qubits, resulting in the entanglement of qubits. a) Schematic figure of protocol. b) Concurrence ($C(\rho_{\rm SS})$) of the two-qubit steady-state of Eq.~(\ref{eqn:2QME}), along with the purity ($\mu$), and logarithmic negativity ($E_{\rm N}$) of the qubits' thermal TMSS environment, all as functions of $r$. The logarithmic negativity is normalized such that its maximum value is one, i.e.~the plot shows $\bar{E}_{\rm N} = E_{\rm N}/E_{\rm N}^{\rm max}$. While $E_{\rm N}$ saturates at a non-zero value, the qubit concurrence has a distinct maximum, eventually dropping to zero as the purity of the thermal TMSS decays. $\bar{\epsilon} = 0.05$ and $\delta=0$ for these curves.}
  \label{fig:QSym}
\end{figure}

The situation changes when there is loss, and the output state from the TWPA becomes a th-TMSS. Consider first the case where the loss is identical for signal and idler modes, and completely frequency independent. As was shown in Ref.~\cite{Grimsmo:2017}, the qubit entanglement (quantified by the concurrence \cite{Wootters:2001aa}) has a distinct maximum as a function of ideal squeezing parameter $r$ (see Fig.~\ref{fig:QSym}), which is at odds with the fact that the logarithmic negativity of a th-TMSS increases monotonically with $r$.  We show here that this can be simply understood as being a result of the decreasing purity of the th-TMSS with increasing $r$.

An example of this is shown in Fig.~\ref{fig:QSym}, where we plot the qubit concurrence in the steady-state, $C(\rho_{\rm SS})$, as a function of the intrinsic gain $G_{\rm ideal} = \cosh^2 r$ of the TWPA.  We also plot the corresponding th-TMSS purity and logarithmic negativity (normalized so that its maximum is one). The qubit concurrence is calculated by solving for the steady-state of the master equation given in Appendix \ref{app:Qubit}. As can be seen, for increasing $r$ the qubit entanglement initially grows as the th-TMSS entanglement; however, very quickly the th-TMSS becomes too impure, and the qubit entanglement rapidly decays.

This qubit-based example highlights the fact that the logarithmic negativity alone is not enough to quantify the usefulness of the entanglement found in a th-TMSS, and therefore from the output of a lossy TWPA.  The purity of the state also plays a crucial role.

\begin{figure}[t]
  \includegraphics[width=0.75\columnwidth]{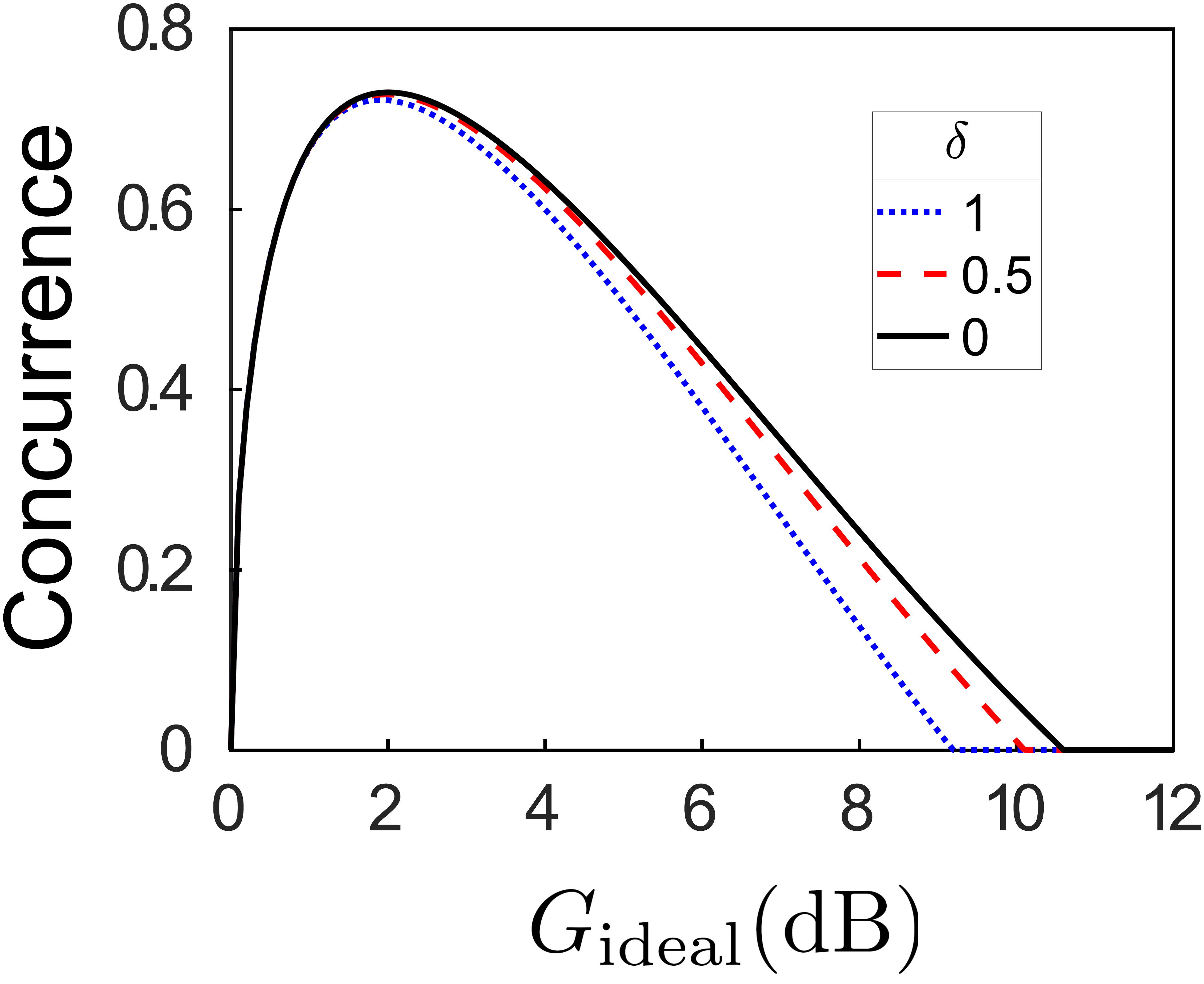}
  \caption{Concurrence ($C(\rho_{\rm SS})$) of the two-qubit steady-state of Eq.~, as a function of $r$, for various values of loss asymmetry $\delta$, with $\bar{\epsilon} = 0.05$ for all three curves. While an asymmetric th-TMSS leads to lower qubit entanglement the difference is minimal, and the asymmetric states perform almost equally as well as the symmetric state in producing qubit entanglement.}
  \label{fig:QAsym}
\end{figure}

\subsubsection{Asymmetric loss}

Extending to the asymmetric case, $\delta \neq 0$, we use the small $\bar{\epsilon}$ expansion for the three th-TMSS parameters derived in Sec.~\ref{sec:th-TMSS} to calculate asymptotic forms of the logarithmic negativity and purity. To second order in $\bar{\epsilon}$ this gives a logarithmic negativity
\begin{align}\label{eqn:lognegexp}
  E_{\rm N}\approx - \ln\left[e^{-2r}+(1-e^{-2r})\bar{\epsilon} +\tanh(r)\bar{\epsilon}^{2}\delta^{2} \right].
\end{align}
While introducing asymmetry further decreases the logarithmic negativity, in the large $r$ limit asymmetry adds only a constant correction of order $\bar{\epsilon}^{2}$ to the expression, and hence does not affect the th-TMSS entanglement significantly.

The purity, similarly to the logarithmic negativity, is also only minimally affected by asymmetry. In the large $r$ limit, the expansion takes the form
\begin{align}
  \mu\approx \frac{1}{1+2(1-\bar{\epsilon})\bar{\epsilon}(\cosh(2r)-1)}-\left(\frac{\bar{\epsilon} \delta}{2(1-\bar{\epsilon})\bar{\epsilon}}\right)^2 e^{-2r}.
\end{align}
As before, the asymmetric correction can be thought of as a renormalization of the coefficient of the exponential decay (since the first term also decays as $\exp(-2r)$ in the large-$r$ limit when $\bar{\epsilon}>0$).

As neither the logarithmic negativity or purity are affected drastically by loss asymmetry it is likely that an asymmetric version of the two-qubit entanglement protocol considered previously will also be only minimally affected. The concurrence of the two-qubit steady-state is shown in Fig.~\ref{fig:QAsym} for various amounts of loss asymmetry, and as can be seen, the fully asymmetric case is only marginally worse than the symmetric case. Thus the asymmetric state is almost as useful as the symmetric state for generating two-qubit entanglement.


\subsection{Correcting for asymmetric loss}\label{sec:correction}

We now consider how to correct for asymmetric loss such that we are able to achieve squeezing below vacuum of {\it symmetric} and commuting joint quadratures, while ideally minimally affecting the purity and logarithmic negativity.  The idea here is simple:  to counteract asymmetric loss, simply add extra loss to the less lossy channel (i.e.~the mode with the larger transmission rate in its effective beam-splitter). Recall that wihtout loss of generality we take $\eta_S<\eta_I$ (see Eq.~(\ref{eqn:eta})), implying the signal mode has more loss than the idler.  Our correction thus corresponds to adding an additional beamsplitter at the idler port with transmission rate $\eta_I' < 1$ (thus further attenuating the idler output). Choosing $\eta'_I =\eta_S/\eta_I$ results in a noise spectral density of the final output $X_{-}$ quadrature given by
\begin{align}
	S^{\rm cor}_{\hat{X}_{-}} = \frac{1}{2}\left[1-\eta_S + \eta_S e^{-2r}\right],
\end{align}
which for $r>0$ is below zero point for any $\eta_S > 0$. Remarkably, by deliberately attenuating the output of the lossy TWPA by a specific amount, we have regained the possibility of squeezing below zero point in a symmetric joint quadrature, as is shown in Fig.~\ref{fig:combinedplot}. Thus, the Tan and Duan inequalities can now be violated while using symmetric, commuting joint quadratures.  Further, the purity and logarithmic negativity of the final output state are minimally affected, taking the value of that for symmetric loss with transmission rate $\eta_S$. Overall, this deliberate introduction of loss can increase the usefulness of the lumped element model lossy TWPA output state in certain applications that require two-mode squeezing of commuting joint quadratures.



\section{Distributed Loss}
\label{sec:dist}

In the beam-splitter model of loss, all losses occur at the end of the amplification region. We now consider a more realistic model of an imperfect TWPA, where loss occurs continuously as photons propagate through the device, and where there is imperfect phase matching between pump, signal and idler.
We start again with the set-up shown in Fig. \ref{fig:setup}(b), as described by Eq.~\eqref{eqn:HamDL}. We now allow both signal idler modes to decay, at independent rates, as they interact parametrically in the region between $x=0$ to $x=L$.

To implement this distributed loss, we follow the approach used in Ref.~\cite{Caves:87} to model loss in a degenerate TWPA.
We imagine connecting independent loss ports at a set of regularly spaced points $x_{j}$ along the TWPA; at each point, there is an independent loss channel for signal and for idler photons.  These loss ports both provide a means for photons to leave the TWPA, and also inject additional vacuum noise into signal and idler modes; their effects are described by standard input-output theory.   We label the injected vacuum noise from these ports as $\hat{a}^{\rm (loss)}_{S/I}(x)$ (where $x$ labels the noise injected at position $x$).  The coupling rate to the loss port at each point is taken to be the same:  $\kappa_S$ for signal photons, $\kappa_I$ for idler photons.
Finally we consider the limit where the spacing between the coupling points $x_j$ tends to zero, resulting in a continuous loss per unit length \cite{Caves:87}.

In addition to this distributed loss, we also now include the effects of imperfect phase matching between pump, signal and idler modes.  Such phase matching is known to be important in realistic TWPAs constructed using Josephson junctions (c.f.~Ref~\cite{Macklin:2015aa,OBrien:2014aa}).  In such system the group velocity is the same for all modes (as they correspond to the same transmission line), but a phase mismatch can arise from nonlinearity-induced frequency shifts.
Imperfect phase matching is characterized by a non-zero wavevector mismatch $\Delta k= 2k_{P}-k_{S}-k_{I}$. For non-zero mismatch, the effective parametric drive in the Hamiltonian has a position dependent phase, c.f.~Eq.(\ref{eqn:parampint}), which serves to disrupt amplification.

With both of these imperfections included, the Heisenberg-Langevin equations for our system become:
\begin{align}\label{eqn:eomDL}
  \left(\partial_{t}+v\partial_{x}  +\frac{i\Delta k}{2}\right)\hat{a}_{S}(x)&=\nu\hat{a}^{\dagger}_{I}(x)-\frac{\kappa_{S}}{2}\hat{a}_{S}(x)\nonumber\\&+\sqrt{\kappa_{S}}\hat{a}^{\rm (loss)}_{S}(x),\\
  \left(\partial_{t}+v\partial_{x}  -\frac{i\Delta k}{2}\right)\hat{a}^{\dagger}_{I}(x)&=\nu\hat{a}_{S}(x)-\frac{\kappa_{I}}{2}\hat{a}^{\dagger}_{I}(x)\nonumber\\&+\sqrt{\kappa_{I}}\hat{a}^{\dagger\rm{ (loss)}}_{I}(x).
\end{align}
Here $v$ is the group velocity (taken to be the same for signal and idler), $\nu$ is the parametric interaction strength, and $\kappa_{S/I}$ are the respective decay rates for the signal/idler modes. These equations are valid from $x=0$ to $x=L$.

Solving the equations of motion in frequency space (see Appendix~\ref{app:DLSol}), we are able to relate the modes at the end of the amplification regions to those at the beginning, allowing us to calculate the system's scattering matrix and performance as a non-degenerate parametric amplifier. Note that within the approximations we use here, the gain and output squeezing of the TWPA are completely independent of frequency; see Appendix~\ref{app:DLSol} for more details.


\subsection{Gain}\label{sec:gain}

\subsubsection{Effects of asymmetric loss}

For the case of symmetric, distributed loss (i.e.~$\kappa_S = \kappa_I$), we find the gain is frequency independent and given by
\begin{align}
  G_{\rm sym}&=e^{-\bar{\kappa} L/v}\cosh^{2}{\left(L\nu/v\right)}\nonumber\\
    &\approx  \frac{e^{(2\nu-\bar{\kappa})L/v}}{4}
\end{align}
which is equivalent to the result of Ref.~\cite{Caves:87} obtained for a degenerate parametric amplifier. This result can be mapped to the effective beam-splitter model of loss in Sec.~\ref{sec:LE}, if we take the beamsplitter transmission to be $\eta=e^{-\bar{\kappa} L/v}$.

For the case of asymmetric loss, a simple mapping to the effective beam-splitter model is no longer possible.  Letting $\kappa_{S}=\bar{\kappa}+\epsilon$, $\kappa_{I}=\bar{\kappa}-\epsilon$, and considering the large length limit, we find the gain with asymmetric loss to be
\begin{align}\label{eqn:Gasym}
  G_{\rm asym}\approx \frac{e^{(2\tilde{\nu}-\bar{\kappa})L/v}}{4}\left[1-\frac{\epsilon}{2\tilde{\nu}}     \right]^2,
\end{align}
where
\begin{align}\label{eqn:enhancednu}
  \tilde{\nu}=\sqrt{\nu^2+\left( \frac{\epsilon}{2} \right)^{2}},
\end{align}
plays the role of a renormalized interaction amplitude.

Comparing symmetric and asymmetric loss results, we see that the for fixed average loss $\bar{\kappa}$, introducing loss asymmetry can increase the gain through its exponential dependence on length (outweighing any reduction due to the non-exponential prefactor).
While this might seem surprising, a similar effect occurs in a simple cavity-based non-degenerate parametric amplifier.  Following the results of Ref. \cite{Clerk:2010aa}, the zero-frequency gain for such a system is given by
\begin{align}\label{eqn:cavitygain}
  \sqrt{G}=\frac{Q^{2}+1}{Q^{2}-1} \hspace{5pt};\hspace{5pt} Q=\frac{2\nu}{\sqrt{\kappa_{S}\kappa_{I}}}
  = \frac{2 \nu }{\bar{\kappa}  \sqrt{1- \epsilon^2}},
\end{align}
where $\kappa_{S/I} = \bar{\kappa} \pm \epsilon$ are the damping rates of signal and idler cavities, and $\nu$ is again the parametric interaction amplitude.  Again, keeping $\bar{\kappa}$ fixed and increasing $\epsilon$ increases the gain.

\subsubsection{Phase mismatch}
We now consider the effects of having imperfect phase matching ($\Delta k \neq 0$) in addition to asymmetric, distributed loss.
For small asymmetry and in the large length limit, the gain becomes
\begin{align}
  G\approx e^{-\bar{\kappa}L/v}e^{2L \Re(\tilde{\nu})/v} \left|   1-\frac{\epsilon+i\Delta k}{2\tilde{\nu}}     \right|^{2},
\end{align}
where the effective complex parametric interaction amplitude is defined as
\begin{align}\label{eqn:phaseMMint}
	  \tilde{\nu}=\sqrt{\nu^2+\left( \frac{\epsilon+i\Delta k}{2} \right)^{2}}.
\end{align}

Without asymmetry, and for large enough $\Delta k$, the effective interaction amplitude $\tilde{\nu}$ becomes purely complex and there is no amplification (i.e.~$G$ remains smaller than one) \cite{Grimsmo:2017,OBrien:2014aa}.  The above result suggests that the effective increase in the parametric interaction amplitude brought on by loss asymmetry can be used to partially offset the decrease in gain due to phase-mismatch.

\subsection{Added noise}

We also consider the added noise of our amplifier with distributed loss and imperfect phase matching.  Recall that even in the ideal case, a non-degenerate parametric amplifier must add half a quantum of noise to the input signal. To calculate the added noise, denoted by $S_{\rm added}$, we look at the spectral density of the output signal mode, normalize by the gain and subtract off the input singal contribution to obtain
\begin{align}\label{eqn:Added}
&2\pi 	S_{\rm added}\delta(\omega+\omega')\equiv\nonumber\\
&\frac{\langle \{ a_{S,\rm{out}}[\omega'],a^{\dagger}_{S,\rm{out}}[\omega] \}\rangle}{2G_{\rm asym}}
	-\frac{\langle \{ a_{S,\rm{in}}[\omega'],a^{\dagger}_{S,\rm{in}}[\omega] \}\rangle}{2} .
\end{align}
Including both asymmetric loss and phase mismatch, in the large gain, small asymmetry, and small phase mismatch limit we find that
\begin{align}
  S_{\rm added} \approx
  	\frac{1}{2} +
		\frac{ 1 }{2\nu-\bar{\kappa}}
		\left(
			\bar{\kappa} + \epsilon
			+ \frac{ (\Delta k^2-\epsilon^2)}{4 \nu(2\nu-\kappa)}
			\right),
\end{align}
independent of frequency. We see that loss increases the added noise above the quantum-limited value of $1/2$, and that this extra noise is sensitive to the amount of asymmetry.  The first order in asymmetry term ($\propto \epsilon$) reflects the fact that $\bar{\kappa} + \epsilon$ is the loss of the signal mode.  The additional terms (which involve both the amount of phase mismatch and asymmetry $\epsilon$) reflect the effective modification of the parametric interaction amplitude due to imperfections, c.f.~Eq.~(\ref{eqn:phaseMMint}).


\subsection{Phase-matched symmetric squeezing}
Having analyzed how the gain and added noise are affected by asymmetric loss, we now focus on the squeezing. We begin by considering the symmetric case where the signal and idler modes decay at the same rate (i.e. $\kappa_{S}=\kappa_{I}\equiv \bar{\kappa}$) and the modes are phase-matched ($\Delta k=0$). In this case, we find that the noise in the squeezed output quadrature is given by
\begin{align}
S^{\rm sym}_{\hat{X}_{-}}[\omega] =  \frac{1}{2(\bar{\kappa}+2\nu)}\left(\bar{\kappa} + 2\nu e^{-2L\nu/v}e^{-\bar{\kappa} L/v} \right). \label{eqn:symDL}
\end{align}
Note there is no mixing-in of amplified noise, and that taking the large-length limit is always beneficial, as the squeezing decreases monotonically with $L$, saturating at a value $\bar{\kappa}/(2 (2 \nu + \bar{\kappa}))$ (c.f.~Fig.~\ref{fig:distloss}).

We find that for symmetric amounts of loss in signal and idler modes, the distributed loss model predicts a less severe degradation of squeezing than the lumped element model of Sec.~\ref{sec:LE}. To see this, we first re-express Eq.~(\ref{eqn:symDL}) as
\begin{align}\label{eqn:mapped}
  S^{\rm sym}_{\hat{X}_{-}}[\omega]=\frac{1}{2} \left[ \left( 1-\eta \right)e^{-2r'} +\eta  e^{-2r}     \right],
\end{align}
where
\begin{align}\label{eqn:mapDL2LE}
&\eta = \frac{2\nu}{\bar{\kappa}+2\nu}e^{-\bar{\kappa} L/v}, \\
&e^{-2r'} = \frac{\bar{\kappa}}{\bar{\kappa}+2\nu\left(1-e^{-{\bar\kappa} L/v}\right)},
\end{align}
and $r=L\nu/v$. This expression is reminiscent of Eq.~(\ref{eqn:symLE}) for the squeezing in the lumped element model of loss.  It is in fact equivalent to a lumped element loss model where squeezed light (characterized by squeezing parameter $r'>0$) is injected into the dark port of the effective beam-splitter (rather than just vacuum noise). By varying the transmission coefficient (via $\bar{\kappa}$), we are interpolating between two different levels of squeezing rather than one level of squeezing and vacuum (no squeezing). Thus, unlike the symmetric lumped element model, the symmetric distributed model always has squeezing below zero point.

A heuristic understanding of this effect comes from our model of symmetric distributed loss, where we consider vacuum noise entering throughout the TWPA, (i.e. from $x = 0$ to $x = L$). One might naively expect this to be detrimental; however, these fluctuations are themselves squeezed by the TWPA interaction. In fact, by considering a spatially varying decay rate ($\bar{\kappa}=\bar{\kappa}(x)$), we can show that only the fluctuations entering near the end of the TWPA section matter. For such a decay rate, the noise in the squeezed output quadrature takes the form
\begin{align}\label{eqn:spatial}
&2	S^{\rm sym}_{\hat{X}_{-}}[\omega] = e^{-\frac{1}{v}\int_{0}^{L}dx\bar{\kappa}(x)}e^{-2L\nu/v}\nonumber\\&+\frac{1}{v}\int_{0}^{L}dx \hspace{3pt}\bar{\kappa}(x) e^{-\frac{1}{v}\int_{x}^{L}dx'\bar{\kappa}(x')}e^{-2\nu (L-x)/v},
\end{align}
where the second term corresponds to the contribution from vacuum fluctuations injected from the loss ports .
In the large-length limit and assuming $\nu\gg \bar{\kappa}(x)$ $\forall x $, the contribution to the squeezing from the added noise integral is exponentially insensitive to noise close to the input port. In other words, it is only the added noise coming from a small region near the end of the amplification region which affects the squeezing. Hence, by minimizing the decay rate near the end of the TWPA section, one could obtain higher amounts of squeezing.


\subsection{Phase-matched asymmetric squeezing}
Next, we consider the asymmetric loss case where $\kappa_{S}\neq \kappa_{I}$, while keeping the phase-matching condition ($\Delta k=0$). Without loss of generality, we assume $\kappa_{S}>\kappa_{I}$ and define $\kappa_{S} = \bar{\kappa} + \epsilon$ , $\kappa_{I} = \bar{\kappa} - \epsilon$, where $\bar{\kappa}$ is the average loss and $\epsilon$ is the asymmetry. The full expression for $S^{\rm asym}_{\hat{X}_{-}}[\omega]$ is cumbersome, and so we consider the low-asymmetry limit. Expanding in the small parameter $\epsilon/\nu$ and keeping only the lowest order term for each possible component of the noise (constant, squeezed, and amplified), we find
\begin{align}\label{eqn:asymDL}
S^{\rm asym}_{\hat{X}_{-}}[\omega] \approx  &\frac{1}{2(\bar{\kappa}+2\nu)}\left[ \bar{\kappa}+2\nu e^{-2L\tilde{\nu}/v}e^{-\bar{\kappa} L/v}   \right]\nonumber\\ &+ G_{\rm eff}\frac{\epsilon^2 e^{-\bar{\kappa} L/v}}{4\nu(2\nu-\bar{\kappa})},
\end{align}
where
\begin{align}
  G_{\rm eff}=\frac{e^{2L\tilde{\nu}/v}}{4},
\end{align}
is an effective gain parameter and $\tilde{\nu}$ is the renormalized interaction strength given by Eq.~(\ref{eqn:enhancednu}). As can be seen in Eq.~\eqref{eqn:asymDL}, asymmetric distributed loss introduces a component to the noise spectral density which scales like the gain of the TWPA. Similarly to the result of Eq.~\eqref{eqn:outvar} for the lumped element model of asymmetric loss, we find that asymmetric distributed loss introduces amplified noise to the $X_-$ quadrature.

\begin{figure}
  \includegraphics[width=0.85\columnwidth]{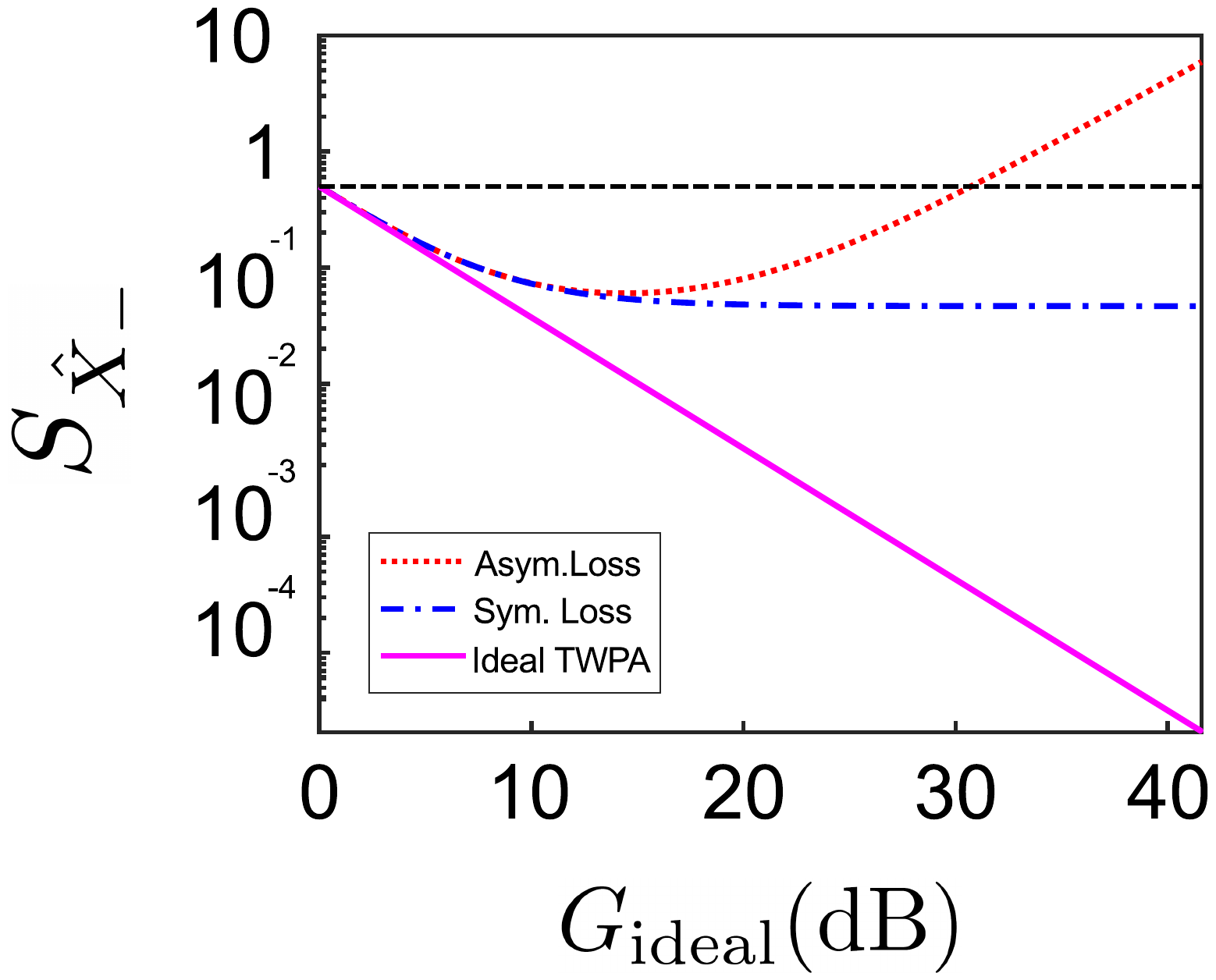}
  \caption{Output squeezing of a TWPA for the distributed loss model (c.f. section \ref{sec:dist}). Curves are plotted as a function of the gain of the ideal distributed model (see Eq.~(\ref{eqn:idealgain})), where the ideal gain is increased by increasing the length $L$. We set $\nu/v=1$, $\bar{\kappa}/v=1/5$, and $\epsilon=\bar{\kappa}/2$. As we increase gain, asymmetric loss (red curve) goes above zero-point squeezing whereas symmetric loss (blue curve) saturates to a value below zero point squeezing. The dashed black line represents zero-point squeezing and the pink curve represents the output squeezing of an ideal TWPA.}
  \label{fig:distloss}
\end{figure}

If the gain is large enough this amplified component dominates the noise, even for small asymmetries. This can be seen in Fig.~\ref{fig:distloss}. The squeezing for asymmetric distributed loss (red curve) initially goes below zero point, however, as the gain for the ideal TWPA increases, our $X_{-}^{\rm out}$ quadrature itself experiences gain. Unlike the symmetric distributed loss case, there is now an optimal length for maximal squeezing below zero point. Working in the low-asymmetry regime, we find that the optimal length is
\begin{align}\label{Lopt}
  L_{\rm opt}\approx \frac{v}{2\nu}\log \frac{\nu}{|\kappa_{S}-\kappa_{I}|}.
\end{align}
Fortunately, we can do better than just using the optimal length to achieve squeezing below zero point in the asymmetric distributed loss case. By correcting for the asymmetry, we can remove this length limitation.


\subsection{Correcting for asymmetric distributed loss}

To correct for the asymmetry, we wish to remove the amplified component of the noise in Eq.~\eqref{eqn:asymDL}. In analogy with the lumped element model, we do so by introducing a beamsplitter on the mode with the smaller decay rate (idler mode), therefore adding additional loss to this mode. Using the full expression for $S^{\rm asym}_{\hat{X}_{-}}[\omega]$, we find that setting the transmission of this beamsplitter to
\begin{align}\label{eqn:mapasym}
\sqrt{\eta_I} = \frac{\nu}{\sqrt{\nu^2+\left(\frac{\epsilon}{2}\right)^2} + \frac{\epsilon}{2}},
\end{align}
completely cancels the coefficient of the amplified component of the noise to all orders. The noise is now given, to lowest order, by the expression
\begin{align}\label{eqn:corrr}
\nonumber S_{\hat{X}_{-}}[\omega] &\approx  \frac{1}{2(\bar{\kappa}+2\nu)}\left(\bar{\kappa} + 2\nu e^{-2L\tilde{\nu}/v}e^{-\bar{\kappa} L/v} \right)  \\ &+ \frac{1}{4}(1-\eta_I).
\end{align}
We see that the corrected low-asymmetry distributed loss squeezing is what would be obtained with symmetric distributed loss (at decay rate $\bar{\kappa}$) plus a constant term coming from the additional beamsplitter loss. Once again, the large-length limit is always beneficial after this correction. Importantly, for sufficiently large gain, adding loss (through the additional beam splitter) allows for squeezing below zero point of the commuting, symmetrically-defined collective quadratures $X_{-}$ and $P_{+}$.


\section{Conclusion}
\label{sec:conc}

In this work, we have studied the effects of frequency-dependent loss on the output state of a TWPA, where photons at signal and idler frequencies see different amounts of dissipation.  Within a simple lumped-element model of loss, we have shown that asymmetric loss can be very detrimental to output squeezing, yet have only minimal effect on the entanglement. It is thus possible to have no joint quadrature squeezing while still having entanglement, and this entanglement may even be useful. By further adding loss to the least lossy mode, we have shown that we are able to regain squeezing below zero point; this could be useful in applications that require the squeezing of symmetric and commuting collective quadratures.

Using a more realistic distributed loss model, we have shown that asymmetric loss increases the gain of the TWPA. By effectively modifying the interaction strength, the exponential dependence of the gain increases. Asymmetric loss can also offset the effects of phase mismatch, to a certain extent, and allow for gain in a situation where it would not occur otherwise. We have shown that when asymmetric loss in included, there is an optimal length for the TWPA after which output squeezing starts to deteriorate. By mapping the distributed loss to a lumped-element model, we show that distributed loss can be thought of as lumped-element loss where we inject squeezed noise rather than vacuum, and that the output squeezing can be corrected in a similar manner as for the true lumped element model.

We note that while we are motivated by the Josephson traveling-wave parametric amplifiers used in circuit QED \cite{Macklin:2015aa}, our results apply universally to traveling-wave non-degenerate parametric amplifiers of any design at any frequency \cite{Kim:1994aa,Boyer:2008aa,Vissers:2016aa,Embrey:2015aa,Erickson:2017,HoEom:2012,Adamyan:2016}. Furthermore, the lumped element model applies to any two-mode squeezing source that is injected into lossy waveguides \cite{Eichler:2011aa,Furst:2011aa,Fortsch:2013aa,Reimer:2016aa}, and our work represents the first exploration of the effects of asymmetric loss in such systems.

We acknowledge useful conversations with Archana Kamal and Simon Gustavsson.  This work was supported by NSERC and the ARO (W911NF-14-1-0078).


\appendix

\section{Master equation for two qubits driven by an imperfect TWPA}
\label{app:Qubit}

The evolution of a pair of qubits sharing a correlated environment, as described in Refs.~\cite{Kraus:2004aa,Grimsmo:2017}, can be described by the master equation (for a detailed derivation consult Ref.~\cite{Grimsmo:2017})
\begin{align}\label{eqn:2QME}
\dot\rho_{q} &= \sum_{k=1,2}\gamma_k\left[\left(1 + N_k\right)\mathcal{D}[\hat{\sigma}_-^k] + N_k\mathcal{D}[\hat{\sigma}_+^k]\right]\rho_q \\ \nonumber&- \sqrt{\gamma_1\gamma_2}M\left(\hat{\sigma}_+^1\rho_q\hat{\sigma}_+^2 + \hat{\sigma}_+^2\rho_q\hat{\sigma}_+^1 - \left\{\hat{\sigma}_+^1\hat{\sigma}_+^2,\rho_q\right\} + h.c. \right)
\end{align}
where $\mathcal{D}[x]\rho = x\rho x^\dagger - \left\{x^\dagger x,\rho_q\right\}/2$ is the usual dissipator, $\hat{\sigma}_{\pm}^k$ are the raising and lowering operators for qubit $k$, and $\gamma_k$ is the coupling rate between qubit $k$ and the environment. The thermal photon population of the environment for each qubit ($N_{k}$), defined by $ \left<\hat{a}_k^\dagger(\omega_k)\hat{a}_k(\omega'_k)\right>~=~2\pi~N_k~\delta(\omega_{k}~+~\omega'_{k})$, as well as the two-qubit anomalous bath correlator (M), defined by $\left<\hat{a}_1(\omega_1)\hat{a}_2(\omega_2)\right>~=~2\pi~M~\delta(\omega_{1}~+~\omega_{2})$, depend on the nature of the environment at the qubit frequencies $\omega_{1/2}$. For the output from a lossy TWPA with signal/idler mode resonant with qubit 1/2 at frequency $\omega/-\omega$, these quantities are given by
\begin{align}
  N_{1/2} &= \bar{n}_{S/I} + (\bar{n}_S + \bar{n}_I+1)\sinh^2(R) = \eta_{S/I}\sinh^2(r),  \\
  M &= \frac{\bar{n}_S + \bar{n}_I +1}{2}\sinh(2R) = \frac{\sqrt{\eta_S\eta_I}}{2}\sinh(2r),
\end{align}
where we have given the form of $N_k$ and $M$ in terms of both the th-TMSS parameterization and the lumped element lossy beamsplitter model. Recall that $\eta_{S}(\eta_{I})$ is evaluated at frequency $\omega(-\omega)$.

For the results of section \ref{sec:PurLN} shown in Figs.~\ref{fig:QSym} and \ref{fig:QAsym}, we solve for the steady-state of Eq.~\eqref{eqn:2QME} numerically, and calculate the concurrence of this state. We set $\gamma_1 = \gamma_2 = \gamma$ for convenience, and in this case the numerical value of $\gamma$ has no effect on the form of the steady-state.

\section{Distributted loss solutions}
\label{app:DLSol}

In this appendix, we provide details on how to obtain the solutions to the distributed-loss model. From the Hamiltonian of Eq.~(\ref{eqn:HamDL}) we obtain the followiing Heisenberg-Langevin equations of motion:
\begin{align}\label{eqn:Ap2}
  \left(\partial_{t}+v\partial_{x}  +\frac{i\Delta k}{2}\right)\hat{a}_{S}(x)&=\nu\hat{a}^{\dagger}_{I}(x)-\frac{\kappa_{S}}{2}\hat{a}_{S}(x)\nonumber\\&+\sqrt{\kappa_{S}}\hat{a}^{\rm (loss)}_{S}(x),\\
  \left(\partial_{t}+v\partial_{x}  -\frac{i\Delta k}{2}\right)\hat{a}^{\dagger}_{I}(x)&=\nu\hat{a}_{S}(x)-\frac{\kappa_{I}}{2}\hat{a}^{\dagger}_{I}(x)\nonumber\\&+\sqrt{\kappa_{I}}\hat{a}^{\dagger \rm (loss)}_{I}(x),
\end{align}
where $\hat{a}^{\rm (loss)}_{S/I}(x)$ is vacuum noise injected at position $x$. To obtain the expressions in this form, we have gauged away the phase of the parametric interaction (recall Eq.~(\ref{eqn:parampint})).

Before tackling the full solution, we begin by solving the differential equations without source terms ($\hat{a}^{\rm (loss)}_{S/I}(x)$). We Fourier transform to frequency space and express everything in matrix form
\begin{align}
  &\partial_{x}\begin{pmatrix}
    \hat{a}_{S}[x,\omega]\\
    \hat{a}^{\dagger}_{I}[x,\omega]
  \end{pmatrix}=\\&\frac{1}{v}\begin{pmatrix}
                            i \omega-(\kappa_{S}+i\Delta k)/2 && \nu\\
                            \nu && i \omega-(\kappa_{I}-i\Delta k)/2
                          \end{pmatrix}\begin{pmatrix}
                                          \hat{a}_{S}[x,\omega]\\
                                          \hat{a}^{\dagger}_{I}[x,\omega]
                                        \end{pmatrix}.\nonumber
\end{align}
The eigenvalues of the matrix on the right-hand side are
\begin{align}
  \lambda_{\pm}=\frac{1}{v}\left(i\omega-\left(  \tfrac{\kappa_{S}+\kappa_{I}}{4}  \right)\pm \sqrt{\nu^{2}+\left(  \tfrac{\kappa_{S}-\kappa_{I}+2i\Delta k}{4}  \right)^{2}}\right),
\end{align}
and the (un-normalized) eigenvectors are
\begin{align}
  \vec{v}_{\pm}=\left(  \tfrac{\kappa_{I}-\kappa_{S}-2i\Delta k}{4\nu} \pm\sqrt{1+\left(  \tfrac{\kappa_{S}-\kappa_{I}+2i\Delta k}{4\nu}  \right)^{2}}       , 1\right)^{T}.
\end{align}
The solutions are given by
\begin{align}
  \begin{pmatrix}
    \hat{a}_{S}[x,\omega]\\
    \hat{a}^{\dagger}_{I}[x,\omega]
  \end{pmatrix}=C_{1}e^{\lambda_{+}x}\vec{v}_{+}+C_{2}e^{\lambda_{-}x}\vec{v}_{-}.
\end{align}
We use the boundary conditions $\hat{a}_{S}[x=0,\omega]=\hat{a}_{S}[0,\omega]$ and $\hat{a}^{\dagger}_{I}[x=0,\omega]=\hat{a}^{\dagger}_{I}[0,\omega]$. We know the signal/idler that enters the chain and we wish to study how they evolve along the TWPA. From these boundary conditions, we can obtain expressions for the coefficients $C_{1}$ and $C_{2}$:
\begin{align}
  C_{1}&=\tfrac{\hat{a}_{S}[0,\omega]-\left(\frac{\kappa_{I}-\kappa_{S}-2i\Delta k}{4\nu} -\sqrt{1+\left(  \frac{\kappa_{S}-\kappa_{I}+2i\Delta k}{4\nu}  \right)^{2}} \right)\hat{a}^{\dagger}_{I}[0,\omega]}{2\sqrt{1+\left(  \frac{\kappa_{S}-\kappa_{I}+2i\Delta k}{4\nu}  \right)^{2}}},\\
  C_{2}&=-\tfrac{\hat{a}_{S}[0,\omega]-\left(\frac{\kappa_{I}-\kappa_{S}-2i\Delta k}{4\nu} +\sqrt{1+\left(  \frac{\kappa_{S}-\kappa_{I}+2i\Delta k}{4\nu}  \right)^{2}} \right)\hat{a}^{\dagger}_{I}[0,\omega]}{2\sqrt{1+\left(  \frac{\kappa_{S}-\kappa_{I}+2i\Delta k}{4\nu}  \right)^{2}}}.
\end{align}
We now wish to express the solutions in the form of a scattering matrix equation
\begin{align}
  \begin{pmatrix}
    \hat{a}_{S}[x,\omega]\\
    \hat{a}^{\dagger}_{I}[x,\omega]
  \end{pmatrix}=\begin{pmatrix}
                            s_{\hat{a}_{S},\hat{a}_{S}}[x,\omega] &&   s_{\hat{a}_{S},\hat{a}^{\dagger}_{I}}[x,\omega]\\
                          s_{\hat{a}^{\dagger}_{I},\hat{a}_{S}}[x,\omega] && s_{\hat{a}^{\dagger}_{I},\hat{a}^{\dagger}_{I}}[x,\omega]
                          \end{pmatrix}\begin{pmatrix}
                                          \hat{a}_{S}[0,\omega]\\
                                          \hat{a}^{\dagger}_{I}[0,\omega]
                                        \end{pmatrix}.
\end{align}
Using the form of $C_{1}$ and $C_{2}$ above, we isolate in terms of $\hat{a}_{S}[0,\omega] $ and $\hat{a}^{\dagger}_{I}[0,\omega]$. The elements of the scattering matrix are
\begin{align}
  s_{\hat{a}_{S},\hat{a}_{S}}[x,\omega] &= e^{(i\omega -(\kappa_{S}+\kappa_{I})/4)x/v} \left[ \cosh(x\tilde{\nu}/v) \right.\nonumber\\ &\left.+\tfrac{\kappa_{I}-\kappa_{S}-2i\Delta k}{4\tilde{\nu}} \sinh(x\tilde{\nu}/v)\right],\nonumber\\
  s_{\hat{a}_{S},\hat{a}^{\dagger}_{I}}[x,\omega] &= s_{\hat{a}^{\dagger}_{I},\hat{a}_{S}}[x,\omega]=e^{(i\omega -(\kappa_{S}+\kappa_{I})/4)x/v}\tfrac{\nu\sinh(x\tilde{\nu}/v)}{\tilde{\nu}},\nonumber\\
  s_{\hat{a}^{\dagger}_{I},\hat{a}^{\dagger}_{I}}[x,\omega] &= e^{(i\omega -(\kappa_{S}+\kappa_{I})/4)x/v} \left[ \cosh(x\tilde{\nu}/v)\right. \nonumber\\ &\left. -\tfrac{\kappa_{I}-\kappa_{S}-2i\Delta k}{4\tilde{\nu}} \sinh(x\tilde{\nu}/v)\right], \label{eqn:Smat}
\end{align}
where
\begin{align}
  \tilde{\nu}=\sqrt{\nu^{2}+\left(  \frac{\kappa_{S}-\kappa_{I}+2i\Delta k}{4}  \right)^{2}}.
\end{align}
We can construct the full solution to the differential equation, including the source terms, using these scattering matrix elements. The full solution is given by
\begin{align}
  \begin{pmatrix}
    \hat{a}_{S}[x,\omega]\\
    \hat{a}^{\dagger}_{I}[x,\omega]
  \end{pmatrix}=&\bf{s}[x,\omega]\begin{pmatrix}
                            \hat{a}_{S}(0)\\
                            \hat{a}^{\dagger}_{I}(0)
                          \end{pmatrix}\nonumber\\ &+\frac{1}{v}\int_{0}^{x}dx'\bf{s}[x-x',\omega]\begin{pmatrix}
                                                    \sqrt{\kappa_{S}}\hat{a}^{\rm (loss)}_{S}[x',\omega]\\
                                                    \sqrt{\kappa_{I}}\hat{a}^{\dagger \rm{(loss)}}_{I}[x',\omega]
                                                  \end{pmatrix}
\end{align}
where $\bf{s}(x)$ is the transfer matrix defined with the above elements in Eq.~\eqref{eqn:Smat}.

\section{Logarithmic negativity and purity}

In this section, we derive the form of Eq.~\eqref{eqn:logneg} from its definition based on the covariance matrix of a two mode squeezed state (taking our two modes to be the signal and idler modes). We define a four-dimensial basis vector $\mathbf{\hat{X}}=(\hat{X}_{S},\hat{P}_{S},\hat{X}_{I},\hat{P}_{I})^{T}$. In this basis, the covariance matrix takes the form
\begin{widetext}
\begin{align}
  \sigma=\begin{pmatrix}
          2\langle \hat{a}^{\dagger}_{S}\hat{a}_{S}  \rangle+1 && 0 && \langle \hat{a}_{I}\hat{a}_{S}  \rangle+\langle \hat{a}^{\dagger}_{I}\hat{a}^{\dagger}_{S}  \rangle && 0 \\
          0 &&   2\langle \hat{a}^{\dagger}_{S}\hat{a}_{S}  \rangle+1 && 0 &&-\langle \hat{a}_{I}\hat{a}_{S}  \rangle-\langle \hat{a}^{\dagger}_{I}\hat{a}^{\dagger}_{S}  \rangle\\
          \langle \hat{a}_{I}\hat{a}_{S}  \rangle+\langle \hat{a}^{\dagger}_{I}\hat{a}^{\dagger}_{S}  \rangle && 0 && 2\langle \hat{a}^{\dagger}_{I}\hat{a}_{I}  \rangle+1 && 0 \\
          0 && -\langle \hat{a}_{I}\hat{a}_{S}  \rangle-\langle \hat{a}^{\dagger}_{I}\hat{a}^{\dagger}_{S}  \rangle && 0 &&2\langle \hat{a}^{\dagger}_{I}\hat{a}_{I}  \rangle+1
  \end{pmatrix}.
\end{align}
\end{widetext}
To find the logarithmic negativity, we need to take the partial transpose of the covariance matrix and then find its eigenvalues. The logarithmic negativty will be given by
\begin{align}
  E_{\rm N}=-\sum_{i} \ln \lambda_{i},
\end{align}
 where $\lambda_{i}$ are distinct eigenvalues with value less than 1. Due to its symplectic form, the partially transposed covariance matrix will only have 2 distinct eigenvalues. Of those two, only one will ever be less than one. We find the eigenvalues to be
 \begin{align}
   \lambda_{\pm}=&\langle \hat{a}^{\dagger}_{S}\hat{a}_{S}  \rangle+\langle \hat{a}^{\dagger}_{I}\hat{a}_{I}  \rangle+1\nonumber\\&\pm\sqrt{\left(\langle \hat{a}^{\dagger}_{S}\hat{a}_{S}  \rangle-\langle \hat{a}^{\dagger}_{I}\hat{a}_{I}  \rangle\right)^{2}+\left(  \langle \hat{a}_{I}\hat{a}_{S}  \rangle+\langle \hat{a}^{\dagger}_{I}\hat{a}^{\dagger}_{S}  \rangle\right)^{2}},
 \end{align}
 where only $\lambda_{-}$ can ever be less than one.

 We can now express the needed averages using the thermal TMSS parameters as introduced in Eqs.~(\ref{eqn:THnum}) and (\ref{eqn:THanom}). A straightforward calculation then yields:
%
%
\begin{align}
	E_{\rm N}=-\ln\left[n_{R}-\sqrt{n_{R}^2-(1+2\bar{n}_{S})(1+2\bar{n}_{I})}       \right].
\end{align}

The purity, as a function of the covariance matrix is given by
\begin{align}
  \mu=\frac{1}{\sqrt{\det(\sigma)}}.
\end{align}
For the case of a TMSS, the eigenvalues of the covariance matrix are the same as the partially transposed one. Since the eigenvalues are repeated, the determinant can be expressed as
\begin{align}
  \det&=\left( \lambda_{+}  \right)^{2}\left( \lambda_{-}  \right)^2\nonumber\\
  &=\left(n^{2}_{R}-\left(\sqrt{n_{R}^2-(1+2\bar{n}_{S})(1+2\bar{n}_{I})} \right)^{2}\right)^{2}\nonumber\\
  &=\left((1+2\bar{n}_{S})(1+2\bar{n}_{I})\right)^{2}.
\end{align}
Hence, the purity takes the form
\begin{align}
  \mu=\frac{1}{(1+2\bar{n}_{S})(1+2\bar{n}_{I})}.
\end{align}

\clearpage

\bibliography{References}

\end{document}